\documentclass[aps,prb,10pt,reprint,superscriptaddress,showpacs]{revtex4-1} 

\usepackage{amsmath}
\usepackage{graphicx}
\usepackage{hyperref}

\usepackage{color}

\begin{document}

\title{Effect of intraband Coulomb repulsion on the excitonic
  spin-density wave}

\author{Bj\"orn Zocher}
\email{zocher@mis.mpg.de}
\affiliation{Max Planck Institute for Mathematics in the Sciences, D-04103
  Leipzig, Germany} 
\affiliation{Institut f\"ur Theoretische Physik, Technische Universit\"at
  Dresden, D-01062 Dresden, Germany} 

\author{Carsten Timm}
\affiliation{Institut f\"ur Theoretische Physik, Technische Universit\"at
  Dresden, D-01062 Dresden, Germany} 

\author{P. M. R. Brydon}
\email{brydon@theory.phy.tu-dresden.de}
\affiliation{Institut f\"ur Theoretische Physik, Technische Universit\"at
  Dresden, D-01062 Dresden, Germany} 

\date{July 08, 2011}

\begin{abstract}
We present a study of the magnetic ground state of a two-band model with nested
electron and hole Fermi surfaces and both interband and
intraband Coulomb interactions. Our aim is to understand how the excitonic
spin-density-wave (ESDW) state induced by the interband Coulomb repulsion is
affected 
by the intraband interactions. We first determine the magnetic instabilities of
our model in an unbiased way by employing the random-phase approximation
(RPA) to calculate the static spin susceptibility in the paramagnetic
state. From this,
we construct the mean-field phase diagram, demonstrating the
robustness of the ESDW against the intraband interaction. We then
calculate the RPA transverse spin susceptibility in the ESDW state and show
that the intraband Coulomb repulsion significantly renormalizes the paramagnon
line shape and suppresses the spin-wave velocity. We conclude with a discussion
of the relevance of this suppression for the commensurate ESDW state of
Mn-doped Cr alloys.   
\end{abstract}

\maketitle

\section{Introduction}
\label{sec:Introduction}

The discovery of superconductivity in the iron pnictides is one of the most
exciting recent developments in condensed matter physics.\cite{KWHH2008}
Although most work has been directed at understanding the
superconducting pairing,~\cite{PG2010} the
unusual antiferromagnetic (AFM) state of the parent compounds has also
attracted much attention.\cite{LC2010} This state appears to be a metallic
spin-density wave 
(SDW), with relatively small staggered magnetic moment at the Fe
sites~\cite{delaCruz2008} and
significant reconstruction of the Fermi surface below the N\'{e}el
temperature $T_{N}$.\cite{Sebastian2008,Yi2009} \emph{Ab initio} calculations
have highlighted the nesting of the 
electron-like and hole-like Fermi surfaces as a crucial ingredient for the
SDW,~\cite{SD2008,MSJD2008} and 
neutron-scattering experiments reveal signatures of itinerant
magnetism.\cite{Ewings2010,Pratt2011}
This has led many theorists to interpret the SDW in the pnictides as a new
manifestation of an old problem: the excitonic instability of a multiband
metal.\cite{HCW2008,KE2008,CEE2008,CT2009,VVC2009,BT2009a,BT2009b,
KEAM2010,KECM2010,FS2010,MC2010,KEM2011} 

The excitonic instability was first proposed in the context of the
semimetal-insulator transition.\cite{C1965,KK1965,KM1965,JRK1967,Z1967}
Assuming electron and hole Fermi pockets separated by a nesting vector ${\bf
  Q}$, the Coulomb repulsion between the two bands can
equivalently be viewed as an attractive interaction between
electrons in one band and holes in the other.
Depending upon the degree of the
nesting, this causes the condensation of interband electron-hole pairs
(excitons) with relative wave vector $\textbf{Q}$ and opens a gap in the
single-particle excitation spectrum.  
Although excitonic semimetal-insulator transitions are rare,~\cite{BF2006} this scenario has
been generalized to account for the presence of additional non-nested Fermi 
surfaces.\cite{R1970} It is widely accepted that such an excitonic
instability is responsible for the metallic SDW state in chromium and its
alloys,~\cite{FM1966,R1970,L1970,B1981,MF1984,F1988,FL1994,FL1996a}  and
the 
excitonic scenario has had notable success in reproducing the spin dynamics
above $T_{N}$ and the doping dependence of the phase diagram.\cite{MF1984}
On the other hand, while it qualitatively captures the spin dynamics
below $T_{N}$, it
nevertheless overestimates the low-temperature spin-wave velocity by a factor
of about 2.\cite{F1988,FL1994,FL1996a} 

The interband interaction responsible for the excitonic instability is only
one of many possible interaction terms for a multiband system. In most
theoretical studies, however,
the intraband interaction is neglected on the basis that it
does not directly play a role in causing interband exciton
formation. The intraband Coulomb repulsion is nevertheless likely to be
at least as  
large as its interband counterpart, and one might expect that it could
give rise to competing magnetic phases or influence the spin
dynamics. These questions are of fundamental interest, since the excitonic spin-density wave (ESDW) is a
key concept in the theory of multiband antiferromagnets. Effective negative, 
i.e., attractive, intraband interactions have been studied in 
Ref. \onlinecite{RKK74}, where they can lead to superconductivity.

In this paper, we present a weak-coupling analysis of
a two-band model with perfect nesting of electron and hole Fermi surfaces and
both interband 
and intraband on-site interactions. We specialize to two dimensions for
consistency with Refs.~\onlinecite{BT2009a,BT2009b},
and also to make contact with the SDW in the iron pnictides.
However, we expect our general results to
be of relevance to any system with nested electron and
hole pockets. After introducing our model in Sec.~\ref{sec:Hamiltonian}, we
start its analysis in Sec.~\ref{sec:Mean Field Theory} by examining the static
spin 
susceptibility in the paramagnetic state, which allows us to determine the
nature of the different magnetic instabilities of the system. This informs a
suitable mean-field ansatz, with which we construct the ground-state phase
diagram of the model. We find that the ESDW state is stable against the
intraband 
interaction at weak to moderate coupling strengths, but is
replaced by states with intraband antiferromagnetic instabilities at
stronger coupling.

In the second part of the paper, we examine the influence of the intraband
interaction on the spin dynamics of the ESDW state.  Although the Dyson
equation for the ESDW state with rather general interband and intraband
interactions has previously been obtained in
Ref.~\onlinecite{BT2009b}, only the interband Coulomb repulsion was assumed
non zero in the numerical evaluation of the transverse spin susceptibility.
In Sec.~\ref{sec:Effect}, we therefore compare the transverse spin
susceptibility  
calculated both with and without accounting for the intraband repulsion.
We show that the finite intraband repulsion leads to a strong renormalization
of the paramagnon line shape and a reduction of the spin-wave velocity. The
relevance of the latter result to the experimental situation in Mn-doped Cr
alloys is discussed in Sec.~\ref{sec:Chromium}, where we argue that the
magnitude of 
the reduction of the spin-wave velocity is consistent with the observed
deviation from the usual weak-coupling predictions. We conclude with a short
summary of our work in Sec.~\ref{sec:Conclusions}.

\section{Model Hamiltonian}
\label{sec:Hamiltonian}

\begin{figure}
\includegraphics[width=\columnwidth]{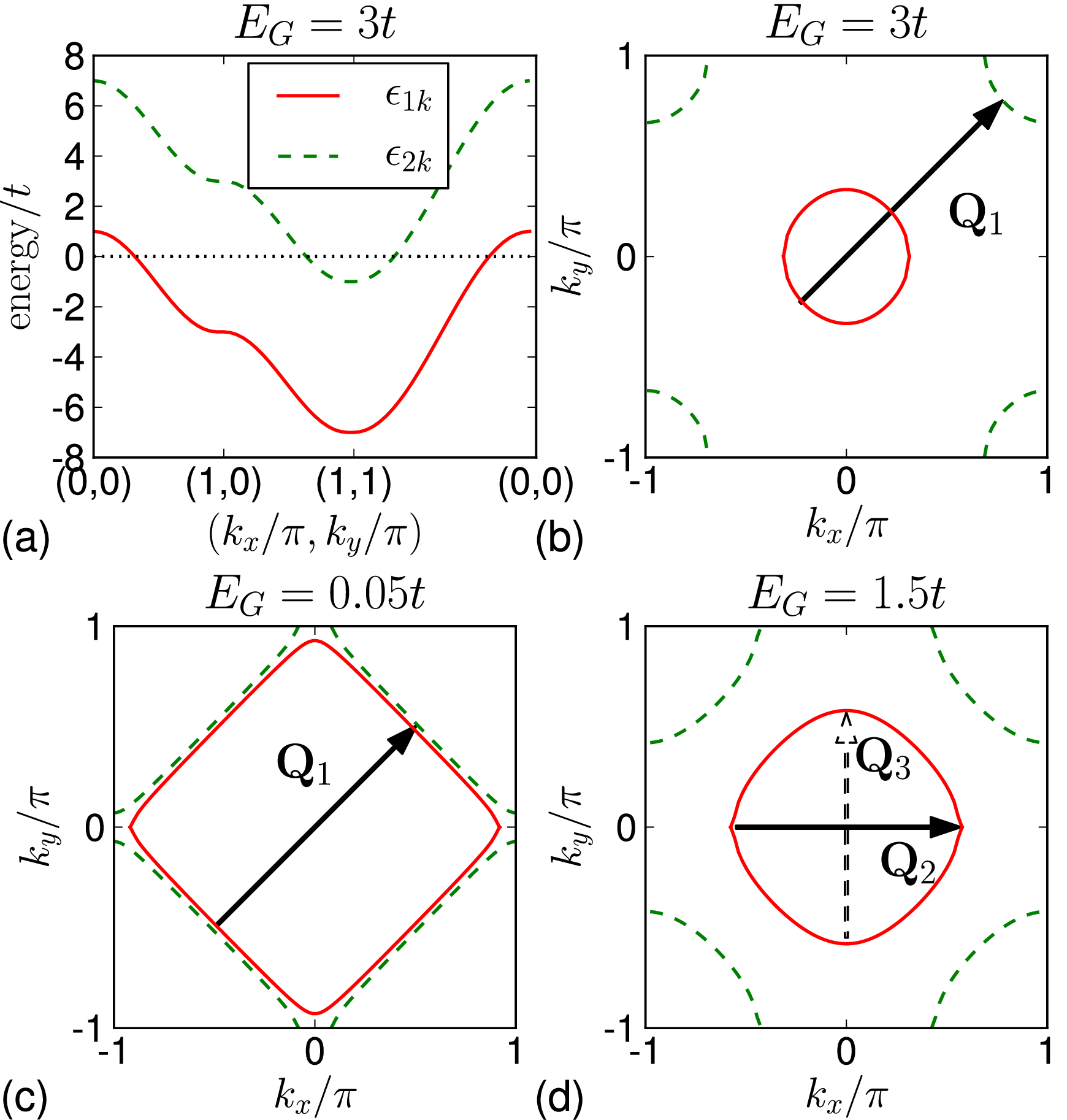}
\caption{(Color online) (a) Band structure and (b) Fermi surface of the
non-interacting model
  for $t_1=t_2=t$ and $E_G = 3\,t$ at half filling. The hole and electron
  pockets are nested by the vector $\textbf{Q}_1=(\pi,\pi)$. The Fermi
  surfaces for $E_G = 0.05\,t$ and $E_G=1.5\,t$ are shown in (c) and (d),
  respectively, illustrating the weaker intraband nesting of parts of the
  electron (hole) Fermi pockets 
  with the vectors $\textbf{Q}_1=(\pi,\pi)$, $\textbf{Q}_2=(\pi,0)$, and
  $\textbf{Q}_3=(0,\pi)$.} 
\label{fig:dispersion_fermisurface_t1_eq_t2}
\end{figure}

We write the minimal Hamiltonian for a two-band semimetal with nested electron
and hole Fermi surfaces as 
\begin{equation}
H = H_0 + H_U +H_I \, .
\label{eqn:H}
\end{equation}
The non interacting Hamiltonian is
\begin{equation}
H_0 = \sum_{\textbf{k},\sigma}\Big[
  \epsilon_{1\textbf{k}}^{}c_{1\textbf{k}\sigma}^\dagger
  c_{1\textbf{k}\sigma}^{}  +   \epsilon_{2\textbf{k}}^{}
  c_{2\textbf{k}\sigma}^\dagger c_{2\textbf{k}\sigma}^{}  \Big] \, , 
\label{eqn:H0}
\end{equation}
where the operator $c_{a \textbf{k} \sigma}^\dagger$ $ (c_{a \textbf{k}
\sigma}^{})$
creates (annihilates) an electron in band $a=1,2$ with momentum
$\textbf{k}$
and spin $\sigma$. For the single-particle energies,  
we consider a two-dimensional band structure with nearest-neighbor hopping, 
\begin{equation}
\epsilon_{ a \textbf{k} } = 2t_a (\cos k_x + \cos k_y) \mp E_G-\mu.
\label{eqn:tighbinding}
\end{equation}
A typical plot of the band structure is given
in~Fig.~\ref{fig:dispersion_fermisurface_t1_eq_t2}(a). 
At half-filling, this band structure always gives a
hole-like pocket at the 
$\Gamma$ point and an electron-like pocket at the $M$ point of the Brillouin
zone. While the parameter $E_G$ tunes the size and shape of the Fermi surface
[see Figs. \ref{fig:dispersion_fermisurface_t1_eq_t2}(b)--\ref{fig:dispersion_fermisurface_t1_eq_t2}(d)],
the electron and hole Fermi pockets are always perfectly nested by the
vector $\textbf{Q}_1=(\pi,\pi)$, i.e., for ${\bf k}$ on the Fermi surface,
we have 
$\epsilon_{1\textbf{k}}=\epsilon_{2\textbf{k}-\textbf{Q}_1}$. 
We note that Eq.~\eqref{eqn:tighbinding} has been
employed as a minimal model of the electronic structure of the iron-pnictide
parent compounds.\cite{HCW2008,VVC2009,CEE2008,BT2009b,MC2010} 

The interaction Hamiltonian consists of three on-site terms which naturally
arise in the effective low-energy theory of multi orbital
models.\cite{CEE2008} Specifically, we have the intraband Coulomb repulsions
within each band,
\begin{equation}
H_U =  \sum_{a=1,2}\frac{U_{aa}}{\mathcal{V}}
 \sum_{\textbf{k},\textbf{k}',\textbf{q}}
 c_{a\textbf{k}+\textbf{q}\uparrow}^\dagger
 c_{a\textbf{k}'-\textbf{q}\downarrow}^\dagger   
 c_{a\textbf{k}'\downarrow}^{}
 c_{a\textbf{k}\uparrow}^{}, 
\label{eqn:HU}
\end{equation}
and the interband Coulomb repulsion,
\begin{equation}
H_I = \frac{U_{12}}{\mathcal{V}}   
 \sum_{\textbf{k},\textbf{k}',\textbf{q}}\sum_{\sigma, \sigma'}  
 c_{1\textbf{k}+\textbf{q}\sigma}^\dagger
 c_{2\textbf{k}'-\textbf{q}\sigma'}^\dagger    
 c_{2\textbf{k}'\sigma'}^{}
 c_{1\textbf{k}\sigma}^{}. 
\label{eqn:HI}
\end{equation}
 For simplicity, we set $U_{22}=U_{11}>0$ in the following, in contrast to 
 previous  theoretical studies where the intraband repulsion is
neglected.\cite{BT2009b,KECM2010,KEM2011}  
The interband Coulomb repulsion is responsible for the 
excitonic instability of the nested electron and hole Fermi surfaces. 
A variety of excitonic mean-field (MF) states are
possible, namely charge-, spin-, charge-current-, and spin-current-density
waves.\cite{B1981,JRK1967,ZFB2010} For the Hamiltonian~Eq.~\eqref{eqn:H},
these density-wave states are degenerate, but the ESDW can be 
stabilized by additional
interband correlated-transition terms.\cite{B1981,BT2009b} These terms can be
assumed to be arbitrarily small, and so we ignore them in our analysis.

\section{Mean Field Theory}
\label{sec:Mean Field Theory}

\subsection{Magnetic instabilities of the paramagnetic state}
\label{sec:Magnetic instabilities}

Within the paramagnetic (PM) state, we obtain an effective mean-field
Hamiltonian by decoupling the interaction terms in~Eq.~\eqref{eqn:H} using
the particle densities
$n_{a\sigma}=1/\mathcal{V}\sum_{\textbf{k}}\langle c_{a \textbf{k}\sigma}
^\dagger c_{a \textbf{k}\sigma}^{} \rangle $. We hence find
\begin{align}
H^{\text{PM}} =& \sum_{a=1,2}  \sum_{\textbf{k},\sigma}   
 \Big(   \epsilon_{a\textbf{k}} + U_{aa} n_{a \bar{\sigma}} +
 U_{12} \sum_{s} n_{\bar{a}s} \Big)
 c_{a\textbf{k}\sigma}^\dagger c_{a \textbf{k}\sigma}^{} \nonumber\\
&-\sum_{a=1,2} U_{aa} \mathcal{V} n_{a\uparrow}n_{a\downarrow}  - 
 U_{12}\mathcal{V}\sum_{\sigma,\sigma'}n_{1\sigma}n_{2\sigma'},
\label{eqn:HPM}
\end{align}
where we introduce the notation $\bar{a}=2(1)$ when $a=1(2)$. Although we
always have perfect nesting, the Hartree terms
in~Eq.~\eqref{eqn:HPM} shift the bands relative to one another, thus changing
the shape of the Fermi surfaces. It is clear from
Figs.~\ref{fig:dispersion_fermisurface_t1_eq_t2}(b)--\ref{fig:dispersion_fermisurface_t1_eq_t2}(d)
 that the changed shape
of the Fermi surface may lead to competing
magnetic phases. These magnetic instabilities can be determined in an
unbiased way by examining the peaks in the PM static spin
susceptibility: as the temperature is
lowered toward the critical temperature of the magnetic state, the static
PM spin susceptibility diverges at the ordering vector $\textbf{Q}$. 

The dynamical spin susceptibility is defined by
\begin{equation}
\chi_{ij,\textbf{q},\textbf{q}'}(i\omega_n) =
 \frac{1}{\mathcal{V}}\int^\beta_0 d\tau \left\langle   T_\tau
 S^i_\textbf{q}(\tau) S^j_{-\textbf{q}'}(0)   \right\rangle e^{i\omega_n\tau}, 
\label{eqn:chipm1}
\end{equation}
where $S^{j}(\textbf{q})$ is the spin operator,
\begin{equation}
S^j_\textbf{q} = \sum_{a,b}   S^{j}_{a,b,\textbf{q}} = \frac{1}{\mathcal{V}}
 \sum_{a,b}  \sum_\textbf{k} \sum_{s, s'}  c_{a\textbf{k}+\textbf{q} s}^\dagger
 \frac{\sigma^j_{s s'}}{2}   c_{b \textbf{k} s'}^{}. 
\label{eqn:sg1}
\end{equation}
Inserting Eq. \eqref{eqn:sg1} into~Eq.~\eqref{eqn:chipm1}, we express the spin
susceptibility in terms of the generalized susceptibilities, 
\begin{eqnarray}
\chi_{ij,\textbf{q},\textbf{q}'}(i\omega_n) & = &
 \sum_{a,b,c,d}\chi_{ij,\textbf{q},\textbf{q}'}^{abcd}(i\omega_n) , \\
 \chi_{ij,\textbf{q},\textbf{q}'}^{abcd}(i\omega_n) &=& \frac{1}{\mathcal{V}}
 \int^{\beta}_{0}d\tau \left\langle   T_{\tau}   S^i_{a,b,\textbf{q}}(\tau)
 S^j_{c,d,-\textbf{q}'}(0) \right\rangle e^{i\omega_n\tau} .\notag \\
&&
\end{eqnarray}
We obtain the static 
transverse MF susceptibilities by making the analytical continuation
$i\omega_n \rightarrow \omega + i0^{+}$ and then 
taking the limit $\omega \rightarrow 0$. 

Due to the invariance of the PM state
under spin rotation, we need only determine the singularities of the
transverse spin susceptibility as these contain all information about the
possible order in the system. By summing up the ladder diagrams, we obtain the
Dyson equation for the generalized RPA spin susceptibilities,
\begin{equation}
\chi^{abba}_{-+,\textbf{q},\textbf{q}}
 =\chi^{abba(0)}_{-+,\textbf{q},\textbf{q}}   +
 U_{ab}\chi^{abba(0)}_{-+,\textbf{q},\textbf{q}}
 \chi^{abba}_{-+,\textbf{q},\textbf{q}} . \label{eqn:PMDyson}
\end{equation}
All other generalized susceptibilities vanish. Expressions for the
lowest-order susceptibilities
$\chi^{abba(0)}_{-+,\textbf{q},\textbf{q}}$ are found in
Ref.~\onlinecite{BT2009b}.
Note that~Eq.~\eqref{eqn:PMDyson} separates into equations for the interband
($a\neq{b}$) and intraband ($a=b$) spin susceptibilities. 

\begin{figure}
\includegraphics[width=\columnwidth]{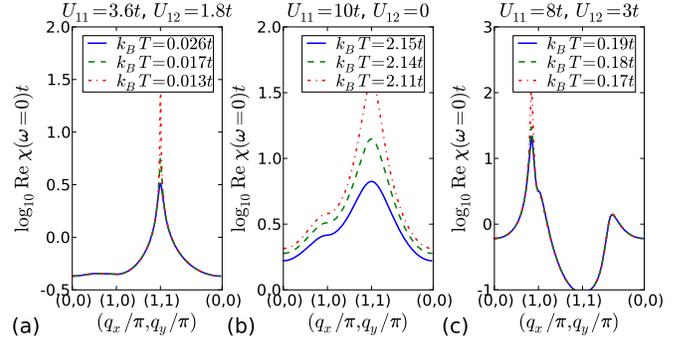}
\caption{(Color online) Total static transverse spin susceptibility for
  three 
  representative points of the parameter space for finite temperatures and
  $t_1=t_2=t$ and $E_G=3t$. Note the different logarithmic scales.} 
\label{fig:static_intraband_susceptibility}
\end{figure}

Evaluating the PM spin susceptibility on a
$2000\times2000$ $\textbf{k}$-point mesh, we find three distinct
magnetic instabilities, which we classify by their ordering vector and 
interband or intraband character.\\
\indent (i) Interband (excitonic) instability with the ordering vector
  $\textbf{Q}=\textbf{Q}_1$, corresponding to the nesting shown in
  Fig.~\ref{fig:dispersion_fermisurface_t1_eq_t2}(b). The evolution of the PM
  susceptibility
  is shown in Fig.~\ref{fig:static_intraband_susceptibility}(a). We describe
  this phase by the order parameter
  $\Delta_{\sigma\sigma'}=(U_{12}/\mathcal{V})\sum_{\textbf{k}}\langle c_{1
    \textbf{k} \sigma}^\dagger c_{2 \textbf{k} - \textbf{Q}_1 \sigma'
  }^{}\rangle$. \\
\indent (ii) Intraband instability with the ordering vector $\textbf{Q}=\textbf{Q}_1$,
  corresponding to the nesting shown in
  Fig.~\ref{fig:dispersion_fermisurface_t1_eq_t2}(c), and the PM
  susceptibilities in Fig.~\ref{fig:static_intraband_susceptibility}(b). We
  describe this instability by the order parameter
  $A_{a\sigma\sigma'}^{(1)}=(U_{11}/\mathcal{V})\sum_{\textbf{k}}\langle c_{a
    \textbf{k} \sigma}^\dagger c_{a \textbf{k} - \textbf{Q}_1 \sigma'
  }^{}\rangle$ with $a=1,2$. \\
\indent (iii) Intraband instability with the ordering vector
  $\textbf{Q}=(\alpha,\beta)$, where $\alpha \approx 0$ and $\beta
  \approx \pi$ or {\it vice versa}, corresponding to the nesting shown in
  Fig.~\ref{fig:dispersion_fermisurface_t1_eq_t2}(d). To describe this
  incommensurate (IC) magnetic order we approximate the vectors by
  $\textbf{Q}_2=(\pi,0)$ and $\textbf{Q}_3=(0,\pi)$. Typical PM
  susceptibilities are shown in
  Fig.~\ref{fig:static_intraband_susceptibility}(c), and we define the order
  parameters 
  $A_{a\sigma\sigma'}^{(\lambda)}=(U_{11}/\mathcal{V})\sum_{\textbf{k}}\langle
  c_{a \textbf{k} \sigma}^\dagger c_{a \textbf{k} -\textbf{Q}_\lambda \sigma'
  }^{}\rangle$ with $a=1,2$ and $\lambda=2,3$.

\subsection{Mean-field phase diagram}
\label{sec:MF Phase Diagram}

We use the order parameters introduced above and the particle densities
$n_{a\sigma}$ to
decouple the interaction terms $H_U$ and $H_I$. Employing standard techniques,
we construct the ground-state MF phase diagram, again using a $2000\times2000$
$\textbf{k}$-point mesh.

\begin{figure}
\includegraphics[width=\columnwidth]{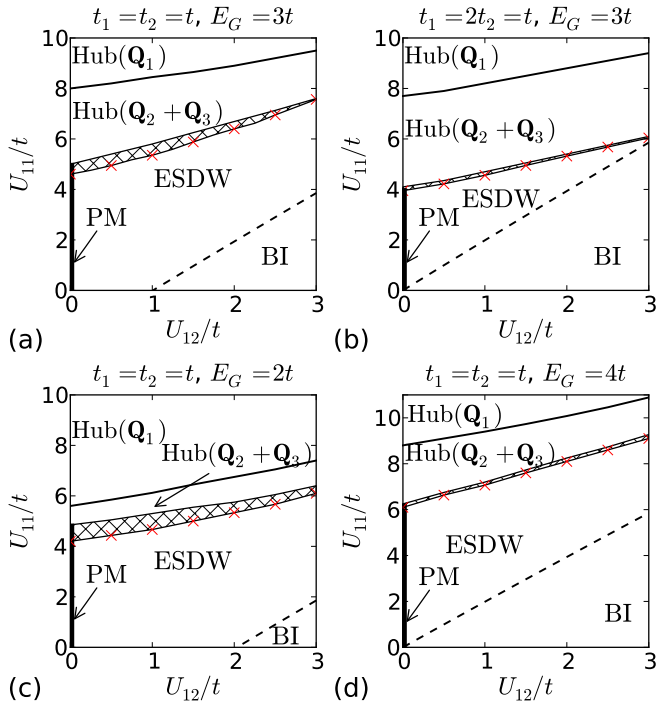}
\caption{(Color online) Ground-state MF phase diagrams for different
  parameters of the band 
  structure at half-filling. 
Solid and dashed lines indicate
  first-order and second-order phase transitions, respectively. Note that the
  transition between the $U_{12}=0$ PM and the ESDW phases is of second
  order. In the shaded region, the MF ground state is the ESDW state, but the
  static spin susceptibility shows an IC AFM intraband instability above the
  critical temperature of the ESDW state. The 
  red crosses show points where the critical
  temperatures of the ESDW and IC AFM states are equal. } 
\label{fig:phase_diagram}
\end{figure}

Figure \ref{fig:phase_diagram} shows four ground-state phase diagrams with
different values of $E_G$ and $t_2$. The structure of these phase diagrams
is quite similar, implying that the topology of the phase diagram is robust
against changes of the band structure. Because of this robustness, we focus on
the plot with $t_1=t_2=t$ 
and $E_G=3\,t$ [Fig.~\ref{fig:phase_diagram}(a)]. We find five different phases:
the PM phase, the 
band-insulator (BI) phase, the 
ESDW phase, the $(\pi,0)+(0,\pi)$ Hubbard AFM phase
[$\mathrm{Hub}(\textbf{Q}_2+\textbf{Q}_3)$], and 
the $(\pi,\pi)$ Hubbard AFM phase [$\mathrm{Hub}(\textbf{Q}_1)$].

We first consider the phase diagram for weak to moderate $U_{11}\lesssim
5\,t$. At 
$U_{12}=0$, we find the PM state. Due to
the perfect nesting of the electron and hole Fermi pockets, however, only an
infinitesimally small $U_{12}$ is required to stabilize the ESDW
phase. The Fermi surface is completely gapped, and we have an insulating
state. Without loss of generality, we take the SDW polarization to be along
the $z$ axis, and so we have the order parameter $\Delta_{\sigma,\sigma'} =
\sigma\delta_{\sigma,\sigma'}\Delta$. Upon increasing $U_{12}$, the 
Hartree shifts $U_{12}n_{\bar{a}}$ in~Eq.~\eqref{eqn:HPM} push the bands
further apart: slightly after the disappearance of the $T=0$ Fermi surface,
the ESDW becomes unstable toward the non magnetic BI phase with  
a completely filled valence and empty conduction band.\cite{I2008}

Starting in the ESDW phase and increasing $U_{11}$, the Hartree shifts $U_{11}
n_{a\bar{\sigma}}$ favor the increase of the occupation of the conduction
band, expanding the electron and hole pockets. For
$n_{1\sigma} \approx 0.78$, the system undergoes a first-order phase 
transition into the $\mathrm{Hub}(\textbf{Q}_2+\textbf{Q}_3)$ state with
finite order parameters
$|A_{a\sigma\sigma'}^{(2)}|=|A_{a\sigma\sigma'}^{(3)}|\neq0$
and arbitrary relative sign.
This phase is a
superposition of magnetically ordered states with intraband ordering vectors
$(\pi,0)$ and $(0,\pi)$, and hence possesses a four-site magnetic unit
cell. At 
higher $U_{11} > 8t$, the system undergoes another 
first-order phase transition into the
$\mathrm{Hub}(\textbf{Q}_1)$ state where $A_{a\sigma\sigma'}^{(1)}\neq0$. The
existence of this phase is not unexpected because in the limit of $U_{12}=0$
and $U_{11}\gg t,E_G$, the system
is equivalent to two independent Hubbard models at half-filling.  

The cross-shaded area indicates the part of the ground-state phase diagram
where the restricted MF calculations predict the ESDW phase but the
susceptibilities show an intraband magnetic instability above the critical
temperature at an IC wave vector. The existence of IC phases in
our two-band model is 
consistent with results for the single-band Hubbard model away from
half-filling.\cite{H1985,S1990}  

To summarize, the ESDW state is robust against the intraband
interaction up to moderate values of $U_{11}$. Indeed, at these strengths, the
Hartree shifts due to the intraband interaction support the ESDW by
suppressing the competing BI phase. Magnetic phases mediated by
the intraband interaction only appear for $U_{11}\gtrsim 5\,t$. This is the
first major result of our work.

\section{Transverse Spin Excitations}
\label{sec:Effect}

\begin{figure}
\includegraphics[clip,width=\columnwidth]{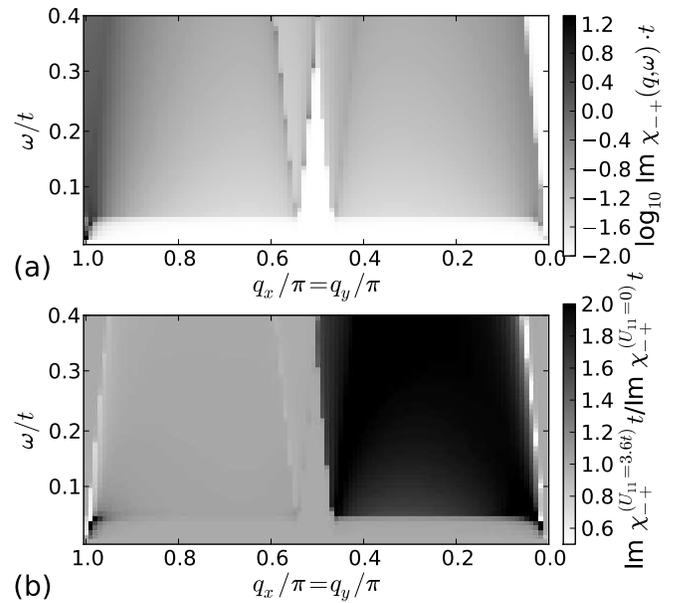}
\caption{(a) Logarithm of the imaginary part of the transverse spin
  susceptibility for $U_{11}=2U_{12}=3.6\,t$ and $\Delta=0.0213\,t$. The spin
wave
  is visible as a dark feature for $\omega < 2\Delta$ near $ \textbf{q}
  =\textbf{Q}_1$ and
  $\textbf{q}=\textbf{0}$. Note the logarithmic color scale. (b)
  Ratio of the total transverse spin susceptibility in panel (a) and the
  $U_{11}=0$ result presented
  in~Ref.~\onlinecite{BT2009b}. }   
\label{fig:logchi}
\end{figure}

The spin excitation spectrum of the ESDW state has unique characteristics
which distinguish it from single-band antiferromagnets.\cite{BT2009b}
We obtain the transverse spin susceptibility for the ESDW state within
the RPA by summing up the ladder diagrams to all orders. This yields the Dyson
equation,
\begin{align}
\chi^{abcd}_{-+ \textbf{q} , \textbf{q}' } =& \delta_{ \textbf{q} ,
  \textbf{q}' } \left( \delta_{a,d}\delta_{b,c}\chi^{abba(0)}_{-+ \textbf{q} ,
  \textbf{q}' }+
\delta_{\bar{a},d}\delta_{\bar{b},c}\chi^{ab\bar{b}\bar{a}(0)}_{-+ \textbf{q}
  , \textbf{q}' }\right) \nonumber \\ 
&+\delta_{ \textbf{q} + \textbf{Q}_1 , \textbf{q}' } \left(
\delta_{\bar{a},d}\delta_{b,c}\chi^{abb\bar{a}(0)}_{-+ \textbf{q} , \textbf{q}'
}+
\delta_{a,d}\delta_{\bar{b},c}\chi^{ab\bar{b}a(0)}_{-+ \textbf{q} ,
  \textbf{q}' }\right) \nonumber \\ 
&+\sum_{\textbf{p}=\textbf{q},\textbf{q}+\textbf{Q}_1}\sum_{m,n=1,2}U_{mn}
\chi^{abmn(0)}_{-+ \textbf{q} , \textbf{p} }\chi^{nmcd}_{-+ \textbf{p} ,
  \textbf{q}' }. \label{eqn:SDWDyson} 
\end{align}
We note that the Dyson equation has been previously obtained in 
Ref.~\onlinecite{BT2009b}, where it was solved only for finite
non zero interband Coulomb repulsion and all other interactions
vanishing.

To calculate the MF transverse spin susceptibilities in
Eq.~\eqref{eqn:SDWDyson}, we used a $10000\times10000$ $\textbf{k}$-point mesh
and a broadening $\delta=10^{-3}t$ in the analytical continuation $i\omega_n
\rightarrow \omega + i\delta$. 
Figure \ref{fig:logchi}(a) shows a typical plot of the imaginary part of the 
transverse spin susceptibility within the ESDW phase for $ \textbf{q} =
(q_x,q_y=q_x)$. We set $t_{1}=t_{2}=t$ and $U_{11}=2U_{12}=3.6\,t$, which gives
a gap
$\Delta=0.0213\,t$. Below, we summarize the main features of the susceptibility;
see 
Ref.~\onlinecite{BT2009b} for a detailed discussion of the susceptibility for 
$U_{11}=0$ and $U_{12}=3.6\,t$. 

As for the static susceptibility calculated in Sec.~\ref{sec:Magnetic
  instabilities}, the total transverse spin 
susceptibility can be divided into contributions 
from intraband and interband excitations. The former gives the response
close to
the zone center, while the latter is responsible for the excitations near
${\bf Q}_1$. The excitation spectrum shows a
partial symmetry of the response about $ \textbf{q} = \textbf{Q}_1 / 2 $. As
shown in Fig. \ref{fig:paramagnon}(a), the distribution of weight also seems
to be a mirror image except for the momenta near $\textbf{Q}_1$ and
$\textbf{0}$. For $\textbf{q}\approx \textbf{0}$, we find a forbidden region
which is anticipated by the considered band structure, while there is a
significant concentration of weight at $\textbf{q}\approx \textbf{Q}_1$.  

The excitation spectrum shows a continuum of single-particle excitations for
$\omega > 2\Delta=0.0426\,t$. This is sharply bounded from below at
$\omega=2\Delta$, which is the minimum energy 
needed to excite quasiparticles across the energy gap of the ESDW state. The
spectrum is bounded by V-shaped features at $ \textbf{q} \approx 0.54 \,
\textbf{Q}_1 $ in the interband susceptibility and at $ \textbf{q} \approx
0.46 \, \textbf{Q}_1 $ in the intraband susceptibility. These features are due
to the weak nesting of parts of the electron Fermi surface with the hole Fermi
surface and with itself, respectively.\cite{BT2009b}
For $\textbf{q} \approx \textbf{Q}_1$, we observe a paramagnon line in the
interband excitation spectrum. 
There is a
similar but much weaker feature in the intraband susceptibility close to
$\textbf{q}=\textbf{0}$.

For $\omega < 2\Delta$, a dispersing spin wave is visible close to the magnetic
ordering vector and, much more weakly, close to the zone center. The
spin-wave dispersion
does not intersect with the single-particle continuum, but instead flattens
out as it approaches $\omega=2\Delta$ and disappears at $\textbf{q} \approx
0.98\, \textbf{Q}_1 $ and $ \textbf{q} \approx 0.02 \, \textbf{Q}_1 $. Although
the paramagnon seems to continue the spin wave into the continuum, closer 
inspection reveals that the two features avoid each other.

\subsection{Effect of the intraband Coulomb repulsion} 

As discussed above, the spin excitation spectrum of the ESDW state is
qualitatively unchanged by the presence of the intraband Coulomb
repulsion. This is unsurprising, as the main features of the transverse
susceptibility are fixed by the ESDW state. It is nevertheless interesting to 
examine the transverse susceptibility for quantitative changes in
experimentally relevant details, such as the paramagnon line shape or the
spin-wave velocity. A direct comparison with the results of
Ref.~\onlinecite{BT2009b} is nevertheless difficult, as Hartree shifts were not
accounted for in that work but instead were assumed to be already included in
$E_G$. This problem can be avoided, however, by choosing $U_{11}=2U_{12}$, for
which the Hartree shifts of the two bands are identical, i.e., effectively
vanishing due to the fixed particle concentration. In this case, the band
structure in the PM state is the same as the one of the non interacting
Hamiltonian. 

\begin{figure}
\includegraphics[width=\columnwidth]{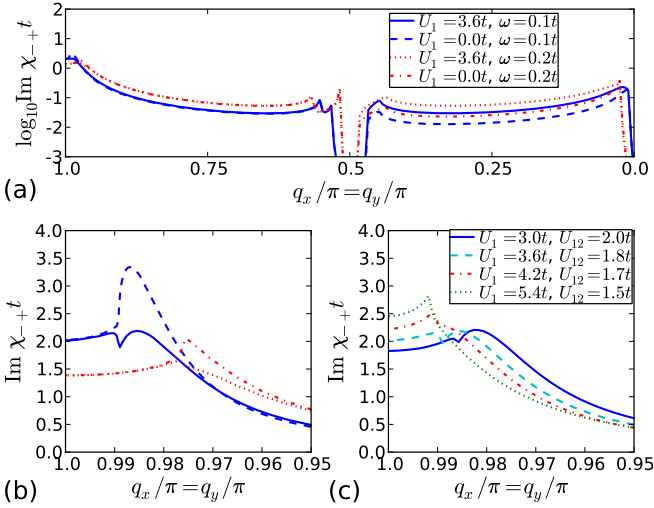}
\caption{(Color online) (a) Cuts through the excitation spectrum for $U_{11}=0$, $U_{12}=1.8\,t$ and $U_{11}=2U_{12}=3.6\,t$. Note the increase of weight of the susceptibility for $q_x < \pi/2$ for $U_{11}\neq 0$, while for $q_x>\pi/2$ the weight changes only close to $q_x=\pi$. (b) Comparison of the paramagnon line shape for the susceptibilities in (a). The lines are defined as in panel (a). (c) Paramagnon line shape for $\omega=0.1\,t$ and various $U_{11}$ along the line of constant $\Delta=0.0213\,t$ in the phase diagram [see Fig. \ref{fig:cs_phase}(b)].} 
\label{fig:paramagnon}
\end{figure}

Figure \ref{fig:logchi}(b) shows the ratio of the transverse spin
susceptibility for $U_{11}=2U_{12} = 3.6\,t$ and
$U_{11}=0$, $U_{12}=1.8\,t$.\cite{BT2009b} As can be seen, the weight
contributed by the intraband components of the spin susceptibility
approximately doubles when we include the intraband 
interaction, while the interband spin susceptibility remains almost the
same. Indeed, as shown in~Fig.~\ref{fig:paramagnon}(a), the intraband and
interband continuum excitations have more nearly equal weight when a finite
$U_{11}$ is present.
Although the interband contribution is not as dramatically affected,
for ${\bf q}\approx{\bf Q}_1$  
the white line at $\omega>2\Delta$ in Fig.~\ref{fig:logchi}(b)
indicates a significant suppression of the
paramagnon by the intraband Coulomb repulsion, while the white-dark feature at
$\omega<2\Delta$ shows a decrease of the spin-wave velocity.

We examine the modification of the paramagnon in greater detail
in Fig.~\ref{fig:paramagnon}(b). At $U_{11}=0$, the paramagnon can be
identified as a distinct peak that becomes broader and lower with increasing
energy $\omega$. 
This changes dramatically at finite $U_{11}$, with the almost complete removal
of the peak. At small excitation energies a ``peak-dip-hump'' structure
develops, while at larger energies the paramagnon looks more like a kink.
Thus, for fixed normal-state band structure, the intraband interaction
can produce a significant change in the paramagnon line shape. In
Fig.~\ref{fig:paramagnon}(c), we show the evolution of
the paramagnon feature with increasing $U_{11}$, where $U_{12}$ is tuned so
that $\Delta$ remains fixed. In contrast to the results in panel (b), here
the normal-state Fermi surface undergoes significant changes due to the
Hartree shifts. In this case, we see that the ``dip-hump'' structure
disappears 
with increasing $U_{11}$, leaving only a progressively sharper peak.
The strong dependence of the paramagnon line shape on
the intraband Coulomb repulsion is the second major result of this paper.

\subsection{Spin-wave velocity}
\label{sec:SW}

\begin{figure}
\includegraphics[width=\columnwidth]{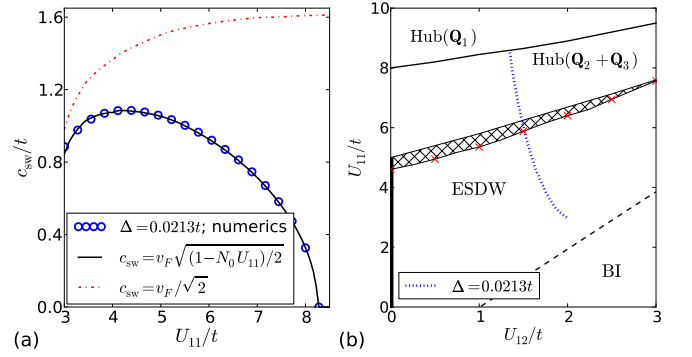}
\caption{(Color online) (a) Comparison of the numerical (open circles) and the approximate (solid and dash-dotted lines) results for the spin-wave velocity for $\Delta=0.0213\,t$ depending on the intraband Coulomb repulsion. (b) Line of $\Delta=0.0213\,t$ (dash-dotted) for which we determine the spin-wave velocity. We continue the constant-$\Delta$ line to the limit of metastability of the ESDW state within the $\mathrm{Hub}(\textbf{Q}_2+\textbf{Q}_3)$ phase. } 
\label{fig:cs_phase}
\end{figure}

The low-energy dispersion of the spin wave can be analytically obtained by
expanding the determinant of the Dyson equation~\eqref{eqn:SDWDyson} about
$\omega = 0$ and ${\bf q} = {\bf Q}_1$. We hence find for the spin-wave
dispersion
\begin{equation}
\omega_\mathrm{m}(\textbf{Q}_1-\textbf{q}) =
c_\mathrm{sw}|\textbf{Q}_1-\textbf{q}|, 
\end{equation}
where $c_\mathrm{sw}$ is the spin-wave velocity. For the band structure
considered here, we have 
\begin{equation}
c_\mathrm{sw}^2 = 2 a_3 \, \frac{(1-2U_{11}a_0 )a_0-2U_{11} \Delta^2
  a_1^2}{a_1^2+2 a_0 a_2 - 8 U_{11} a_0 a_1^2}, 
\label{eqn:spinwave_coff}
\end{equation}
where
\begin{align}
a_0 &= \frac{1}{4\mathcal{V}}  \sum_\textbf{k}
\frac{\Delta^2}{E_\textbf{k}^3},~~ a_1 = \frac{1}{4\mathcal{V}}
\sum_\textbf{k}   \frac{\tilde{\epsilon}_{1 \textbf{k}}}{E_\textbf{k}^3},~~ a_2
= \frac{a_0}{2\Delta^2},\\ 
a_3 &= \frac{t}{2\mathcal{V}}  \sum_\textbf{k}  \left( t \, \frac{2\Delta^2 -
  \tilde{\epsilon}_{ 1 \textbf{k}}^2 }{E_\textbf{k}^5} \,  \mathrm{sin}^2 k_x-
\frac{\tilde{\epsilon}_{ 1 \textbf{k}} }{2E_\textbf{k}^3} \,
\mathrm{cos}k_x\right) . 
\end{align}
Note that the definition of $a_3$ is different from that in
Ref. \onlinecite{BT2009b}. In the physically relevant limit $\Delta \ll t$, we
find 
\begin{equation}
c_\mathrm{sw} \cong \frac{v_F}{\sqrt{2}}\sqrt{1-\mathcal{N}_0 U_{11}},
\label{eqn:apprx_csw}
\end{equation}
where $v_F$ is the average Fermi velocity and $\mathcal{N}_0$ is the single-spin
density of states of one of the bands at the Fermi energy. 
Equation~\eqref{eqn:apprx_csw} is the final major result of this work.
For $U_{11}=0$, we recover the well-known relation $c_\mathrm{sw} \cong v_F /
\sqrt{2}$.\cite{FM1966,L1970,BT2009b} At finite $U_{11}$, the spin-wave
velocity can be significantly suppressed by the
factor $\sqrt{1-\mathcal{N}_0 U_{11}}$, and it
exactly vanishes when the Stoner criterion for (intraband) ferromagnetism
is satisfied. 
The dependence of the spin-wave velocity on $U_{11}$ can be seen in
Fig. \ref{fig:cs_phase}, where we plot the spin-wave velocity for a cut
through the phase diagram at constant order parameter $\Delta$. We find an
excellent agreement between~Eq.~\eqref{eqn:apprx_csw} and the numerical data,
with the spin-wave velocity going through a maximum as $U_{11}$ is
increased. In contrast, the usual expression $c_\mathrm{sw} = v_F
/\sqrt{2}$ overestimates the spin-wave velocity, and monotonically increases
with $U_{11}$ due to the effect of the Hartree shifts upon $v_F$.

\section{Spin-Wave Velocity in Chromium Alloys}
\label{sec:Chromium}

The ESDW state is widely believed to be realized in Cr and its
AFM alloys. Cr displays a slightly incommensurate SDW with a N\'eel
temperature of 311 K and a temperature dependence of the staggered
magnetization that is well described by standard MF theory.\cite{F1988} A
commensurate SDW state with much higher $T_N$ can be stabilized by doping with
Mn.\cite{KMTM1966} Several authors have discussed the spin dynamics of such
Cr alloys using the RPA.\cite{FM1966,L1970,FL1994,FL1996a} Neglecting
intraband interactions, they found good agreement between theory and
experiment above the N\'eel temperature $T_N$.\cite{SL1969,N1971} At low
temperatures $T \ll T_N$, a key theoretical prediction is that the spin-wave
velocity is $c_\mathrm{sw}^2 = v_e v_h /3$, where $v_{h(e)}$ is the hole
(electron) Fermi velocity and the factor of $3$ in the denominator arises
because a three-dimensional band structure is
considered.\cite{FL1994,L1970,FM1966,FL1996a} Experiments, however, show that
this result overestimates $c_{\mathrm{sw}}$ by a factor of approximately
2.\cite{SL1969,N1971,SK1977} 
This discrepancy persists in more sophisticated models of the band structure
(e.g., Ref.~\onlinecite{FL1996a}) and has not yet been conclusively
explained. A notable attempt was made by Liu,~\cite{L1976,L1981} who
proposed that the coupling
between SDW-induced local moments on the Cr ions and magnons was
responsible for the reduced spin-wave velocity.

In Sec.~\ref{sec:Effect}, we found that the intraband Coulomb repulsion
provides a significant renormalization of the spin-wave velocity in the
ESDW phase. It is therefore interesting to estimate the effect of this
renormalization for Mn-doped Cr alloys [the result,
Eq.~\eqref{eqn:apprx_csw}, can easily be generalized to a three-dimensional
system by replacing the factor $\sqrt{2}$ by $\sqrt{3}$]. {\it Ab initio}
calculations estimate the total density of states of paramagnetic Cr to be
approximately
$0.65$ eV$^{-1}$.\cite{S1981,LCFB1981} Ignoring non-nested portions
of the Fermi surface, this gives an upper bound ${\cal N}_{0} \lesssim 0.16$
eV$^{-1}$. Liu\cite{L1970} estimated ${\cal N}_{0}U_{12}=0.43$ by fitting
to experimental
data on Cr$_{0.98}$Mn$_{0.02}$. It is therefore reasonable to take 
$U_{11} \approx 3\,$eV, which gives a renormalization 
$\sqrt{1-\mathcal{N}_0 U_{11}} \approx 0.7$, i.e., the renormalization factor
yields a reduction of the spin-wave velocity by approximately 30\ \%. This
accounts for the bulk of the discrepancy between the $U_{11}=0$ theory and
experimental findings, and suggests that the hitherto neglected intraband
interactions could play a significant role in the spin dynamics of Cr and its
alloys. We note that a finite $U_{11}$ will not affect the low-energy
  normal-state spin dynamics near to the magnetic ordering vector, and so the
  previous results for $T>T_{N}$ are also valid in our theory.

\section{Conclusions} \label{sec:Conclusions}

In this paper, we have presented an analysis of a two-dimensional two-band
Hubbard model with nested electron and hole Fermi surfaces. By examining the
static RPA spin susceptibility in the paramagnetic state, we have determined
the possible magnetic order in an unbiased way. In addition to
the expected interband ESDW, we have found instabilities toward a number of
intraband AFM states with various commensurate and incommensurate ordering
vectors. Using these results to inform a mean-field ansatz, we have
calculated the ground-state phase diagram for a number of different 
semimetallic band structures. We have shown that the ESDW
state is stable at weak to 
moderate intraband coupling; at stronger interaction strengths, however,
changes in the Fermi surface induced by the Hartree shifts stabilize the
intraband AFM states. 

In the second part of the paper, we have studied the effect of the
intraband
interactions on the low-temperature spin dynamics of the ESDW. We
have solved the
Dyson equation for the transverse spin susceptibility and have compared the
results for vanishing~\cite{BT2009b} and finite intraband interactions.
We find that there is significant renormalization of key
experimentally relevant details of the spin excitation spectrum due to
intraband interactions. Specifically, the intraband interactions
qualitatively alter the paramagnon line shape and reduce the spin-wave
velocity. We argue that this mechanism could resolve the discrepancy between 
the measured spin-wave velocity in Mn-doped Cr and previous theoretical 
predictions.

\acknowledgments
The authors thank M. Daghofer, I. Eremin, J. Knolle, J. Schmiedt, J. van
den Brink, and M. Vojta for useful discussions. B.Z. gratefully acknowledges
support by the Studienstiftung des Deutschen Volkes. C.T. and P.M.R.B
acknowledge support from the Deutsche Forschungsgemeinschaft under Priority
Programme 1458.


\begin{thebibliography}{99}

\bibitem{KWHH2008}Y. Kamihara, T. Watanabe, M. Hirano, and H. Hosono,
  J. Am. Chem. Soc. {\bf 130}, 3296 (2008).

\bibitem{PG2010}J. Paglione and R. L. Green, Nature Phys. {\bf 6}, 645 (2010).

\bibitem{LC2010}M. D. Lumsden and A. D. Christianson, J. Phys.:
  Condens. Matter {\bf 22}, 203203 (2010).

\bibitem{delaCruz2008}C. de la Cruz, Q. Huang, J. W. Lynn, J. Li, W. Ratcliff
  II, J. L. Zaretsky, H. A. Mook, G. F. Chen, J. L. Luo, N. L. Wang, and
  P. Dai, Nature (London) {\bf 453}, 899 (2008).

\bibitem{Sebastian2008}S. E. Sebastian, J. Gillett, N. Harrison, P. H. C. Lau,
  C. H. Mielke, and G. G. Lonzarich, J. Phys.: Condens. Matter {\bf 20},
  422203 (2008).

\bibitem{Yi2009}M. Yi, D. H. Lu, J. G. Analytis, J.-H. Chu, S.-K. Mo,
  R.-H. He, M. Hashimoto, R. G. Moore, I. I. Mazin, D. J. Singh, Z. Hussain,
  I. R. Fisher, and Z.-X. Shen, Phys. Rev. B {\bf 80}, 174510 (2009).

\bibitem{SD2008}D. J. Singh and M.-H. Du, Phys. Rev. Lett. {\bf 100}, 237003
  (2008). 

\bibitem{MSJD2008}I. I. Mazin, D. J. Singh, M. D. Johannes, and M. H. Du,
  Phys. Rev. Lett. {\bf 101}, 057003 (2008).

\bibitem{Ewings2010}R. A. Ewings, T. G. Perring, J. Gillett, S. D. Das,
  S. E. Sebastian, A. E. Taylor, T. Guidi, and A. T. Boothroyd,
  Phys. Rev. B {\bf 83}, 214519 (2011).

\bibitem{Pratt2011}D. K. Pratt, M. G. Kim, A. Kreyssig, Y. B. Lee,
  G. S. Tucker, A. Thaler, W. Tian, J. L. Zarestky, S. L. Bud'ko,
  P. C. Canfield, B. N. Harmon, A. I. Goldman, and R. J. McQueeney,
  Phys. Rev. Lett. {\bf 106}, 257001 (2011).

\bibitem{HCW2008}Q. Han, Y. Chen, and Z. D. Wang, Europhys. Lett. {\bf 82},
  37007 (2008).

\bibitem{KE2008}M. M. Korshunov and I. Eremin, Phys. Rev. B {\bf 78},
  140509(R) (2008).

\bibitem{CEE2008}A. V. Chubukov, D. V. Efremov, and I. Eremin, Phys. Rev. B
  {\bf 78}, 134512 (2008).

\bibitem{CT2009}V. Cvetkovic and Z. Tesanovic, Europhys. Lett. {\bf 85}, 37002
  (2009); Phys. Rev. B {\bf 80}, 024512 (2009).

\bibitem{VVC2009}A. B. Vorontsov, M. G. Vavilov, and A. V. Chubukov,
  Phys. Rev. B {\bf 79}, 060508(R) (2009).

\bibitem{BT2009a}P. M. R. Brydon and C. Timm, Phys. Rev. B {\bf 79}, 180504(R)
  (2009). 

\bibitem{BT2009b}P. M. R. Brydon and C. Timm, Phys. Rev. B {\bf 80}, 174401
  (2009). 

\bibitem{KEAM2010}J. Knolle, I. Eremin, A. Akbari, and R. Moessner,
  Phys. Rev. Lett. {\bf 104}, 257001 (2010).

\bibitem{KECM2010}J. Knolle, I. Eremin, A. V. Chubukov, and R. Moessner,
  Phys. Rev. B {\bf 81}, 140506(R) (2010).

\bibitem{FS2010}R. M. Fernandes and J. Schmalian, Phys. Rev. B {\bf 82},
  014521 (2010).

\bibitem{MC2010}S. Maiti and A. V. Chubukov, Phys. Rev. B {\bf 82}, 214515
  (2010). 

\bibitem{KEM2011}J. Knolle, I. Eremin, and R. Moessner, Phys. Rev. B {\bf 83}, 224503 (2011). 

\bibitem{C1965}J. des Cloizeaux, J. Phys. Chem. Solids {\bf 26}, 259 (1965).

\bibitem{KK1965}L. V. Keldysh and Y. V. Kopaev, Fiz. Tverd. Tela (Leningrad) 6, 2791 (1964) [Sov. Phys. Solid State 6, 2219 (1965)].

\bibitem{KM1965}A. N. Kozlov and L. A. Maksimov, Zh. Eksp. Teor. Fiz. 48, 1184 (1965) [Sov. Phys. JETP 21, 790 (1965)]. 

\bibitem{JRK1967}D. Jerome, T. M. Rice, and W. Kohn, Phys. Rev. {\bf 158}, 462
  (1967). 

\bibitem{Z1967}J. Zittartz, Phys. Rev. {\bf 162}, 752 (1967).

\bibitem{BF2006}F. X. Bronold and H. Fehske, Phys. Rev. B {\bf 74}, 165107
  (2006). 

\bibitem{R1970}T. M. Rice, Phys. Rev. B {\bf 2}, 3619 (1970).

\bibitem{FM1966}P. A. Fedders and C. A. Martin, Phys. Rev. {\bf 143}, 245
  (1966).

\bibitem{L1970}S. H. Liu, Phys. Rev. B {\bf 2}, 2664 (1970).

\bibitem{B1981}D. W. Buker, Phys. Rev. B {\bf 24}, 5713 (1981).

\bibitem{MF1984}K. Machida and M. Fujita, Phys. Rev. B {\bf 30}, 5284 (1984).

\bibitem{F1988}E. Fawcett, Rev. Mod. Phys. {\bf 60}, 209 (1988).

\bibitem{FL1994}R. S. Fishman and S. H. Liu, Phys. Rev. B {\bf 50}, 4240(R)
  (1994).

\bibitem{FL1996a}R. S. Fishman and S. H. Liu, Phys. Rev. B {\bf 54}, 7233
  (1996). 
  
\bibitem{RKK74}A. I. Rusinov, D. C. Kat, and Y. V. Kopaev, Zh. Eksp. Teor. Fiz. 65, 1984 (1973) [Sov. Phys. JETP 38, 991 (1974)]; M. Gul\'{a}csi and Zs. Gul\'{a}csi, Phys. Rev. B {\bf 39}, 714 (1989).

\bibitem{ZFB2010}B. Zenker, H. Fehske, and C. D. Batista, Phys. Rev. B {\bf
  82}, 165110 (2010).

\bibitem{I2008}D. Ihle, M. Pfafferott, E. Burovski, F. X. Bronold, and
  H. Fehske, Phys. Rev. B {\bf 78}, 193103 (2008).

\bibitem{H1985}J. E. Hirsch, Phys. Rev. B {\bf 31}, 4403 (1985).

\bibitem{S1990}H. J. Schulz, Phys. Rev. Lett. {\bf 64}, 1445 (1990).

\bibitem{KMTM1966}W. C. Koehler, R. M. Moon, A. L. Trego, and
  A. R. Mackintosh, Phys. Rev. {\bf 151}, 405 (1966).

\bibitem{SL1969}S. K. Sinha, S. H. Liu, L. D. Muhlestein, and N. Wakabayashi,
  Phys. Rev. Lett. {\bf 23}, 311 (1969).

\bibitem{N1971}J. Als-Nielsen, J. D. Axe, and G. Shirane, J. Appl. Phys. {\bf
  42}, 1666 (1971).

\bibitem{SK1977}S. K. Sinha, G. R. Kline, C. Stassis, N. Chesser, and
  N. Wakabayashi, Phys. Rev. B {\bf 15}, 1415 (1977).

\bibitem{L1976}S. H. Liu, Phys. Rev. B {\bf 13}, 3962 (1976).

\bibitem{L1981}S. H. Liu, J. Magn. Magn. Mater. {\bf 25}, 97 (1981).

\bibitem{S1981}H. L. Skriver, J. Phys. F {\bf 11}, 97 (1981).

\bibitem{LCFB1981}D. G. Laurent, J. Callaway, J. L. Fry, and N. E. Brener,
  Phys. Rev. B {\bf 23}, 4977 (1981).

\end{thebibliography}
\end{document}